# Simple Formulas for a Two-Tier Strategy to Flatten the Curve

Michael Nikolaou[1]

[1] Chemical and Biomolecular Engineering Department, University of Houston, Houston, TX 77204

E-mail: nikolaou@uh.edu

*Abstract* – Some basic facts that enable or constrain efforts to "flatten the curve" by non-pharmaceutical interventions (NPI) are summarized and placed in the current context of the COVID-19 epidemic. Analytical formulas are presented for a simple two-tier NPI strategy that places different social distancing targets for high- and low-risk groups. The results aim to facilitate rapid what-if analysis rather than replace detailed simulation, which remains indispensable for informed decision making. The appearance of the oft neglected Lambert function in the resulting formulas is noted.

*Keywords* – COVID-19, Coronavirus, Flattening the Curve, Social Distancing, Population Dynamics, Control

## 1. Introduction

In the absence of effective vaccines or antiviral drugs against COVID-19, non-pharmaceutical interventions (NPI) – measures aiming at reducing virus transmission in the population through reduction of human contacts – remain the only viable response to the epidemic.[3] The inexorable realities faced by NPI are best accentuated by the simple plot in Figure 1 resulting from analysis of the standard SIR epidemiology model: With the basic reproductive ratio, $R_0$, for COVID-19 being between 2 and 3,[2] the total fraction of a population infected by the end of the epidemic – whenever that might come – would be around 90% (for $R_0 \approx 2.5$ in Figure 1) in the absence of NPI. At a fraction of 1% mortality for infected individuals, fatalities for a population the size of the US would be in the millions. This bleak estimate is in agreement with results from more elaborate calculations earlier presented by the US Centers for Disease Control and Prevention (CDC).[4]

NPI measures should bring $R_0$ down from 2.5 to about 1.0 for the infection rate to eventually reverse sign to negative, thus eventually resulting in low single digits for total percentage of infected individuals in a population (Figure 1); or to below 1, for the epidemic to immediately start contracting. For example, it is reported that travel restrictions in Wuhan, China quickly brought $R_0$ down from 2.35 to 1.05.[2] Similarly, recent projections by the US federal government for 200,000 fatalities correspond to $R_0 \approx 1.03$.[5] As a practical matter, current data on exponential growth of infections in various countries (widely available on the internet[6]) indicate that extraordinary measures are needed for $R_0$ to be drastically reduced. With a sizable asymptomatic period for infected individuals, universal restrictions on social contacts quickly emerged as a preferred NPI measure.[1]

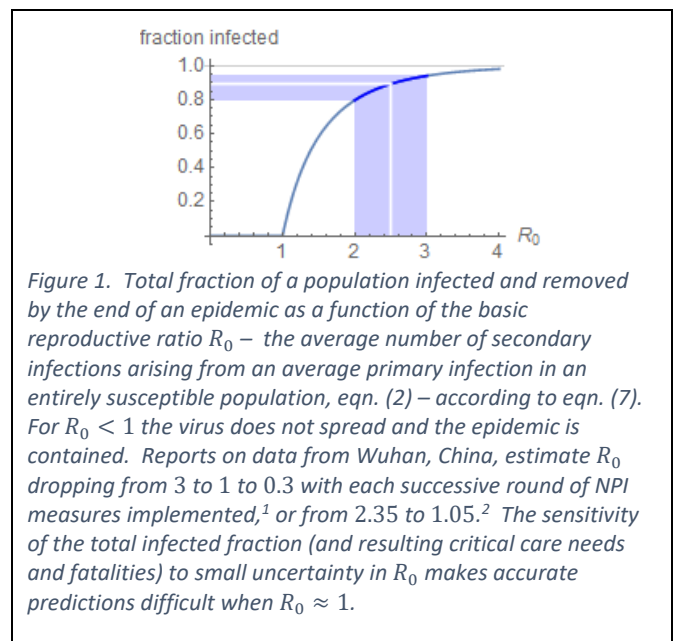

*Figure 1. Total fraction of a population infected and removed by the end of an epidemic as a function of the basic reproductive ratio $R_0$ – the average number of secondary infections arising from an average primary infection in an entirely susceptible population, eqn. (2) – according to eqn. (7). For $R_0 < 1$ the virus does not spread and the epidemic is contained. Reports on data from Wuhan, China, estimate $R_0$ dropping from 3 to 1 to 0.3 with each successive round of NPI measures implemented,[1] or from 2.35 to 1.05.[2] The sensitivity of the total infected fraction (and resulting critical care needs and fatalities) to small uncertainty in $R_0$ makes accurate predictions difficult when $R_0 \approx 1$.*

In addition to the role of $R_0$ already discussed, it turns out that reducing $R_0$ helps mitigate the overwhelming of hospitalization facilities by infected individuals in need of treatment, and thus prevent fatalities from inadequately treated infections. This is because the fraction of infectious individuals reaches its peak value of $1 - 1/R_0 - \ln(R_0)/R_0$ at a time approximately proportional to $1/(R_0 - 1)$ (*Appendix B*). In this context, lowering that peak by reducing $R_0$ achieves "Flattening the Curve,"[6-8] as discussed in more detail in section 3.

These realizations have culminated into the policy "Stay Home", as a practical if universal NPI measure to effect the necessary reduction in human contacts and thus lower $R_0$. However, in addition to flattening the curve, this reduction also has adverse effects felt by many, not least in terms of drastically reduced economic activity. Facing the inexorable conflict summarized as "saving lives vs. saving the economy" appears inevitable. The purpose of this publication is to help understand that conflict and contribute towards its mitigation by exploiting a simple and widely observed fact: That hospitalization and fatality rates are higher by at least an order of magnitude among the elderly and the vulnerable who are roughly 1/5 to 1/6 of the population (Figure 2).[9,10]

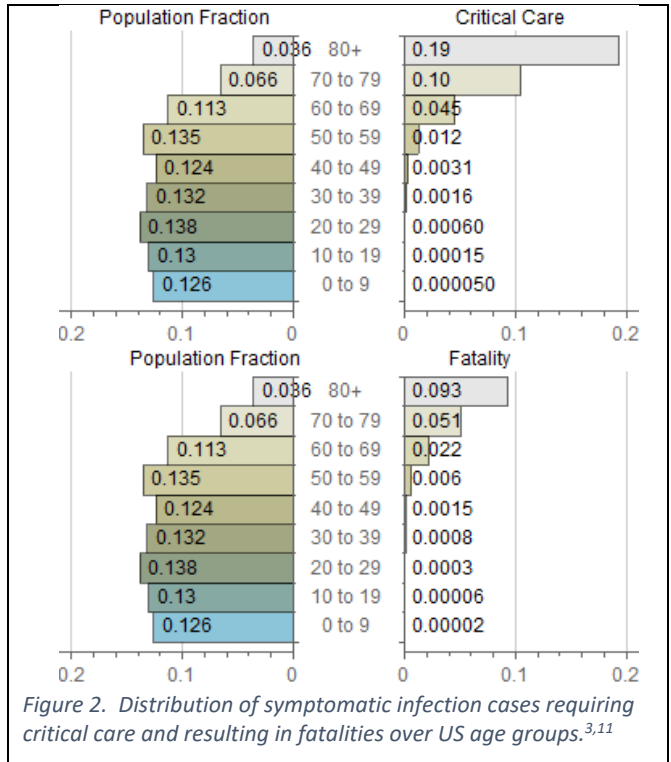

*Figure 2. Distribution of symptomatic infection cases requiring critical care and resulting in fatalities over US age groups.[3,11]*

Therefore, by shielding human contact for this *high-risk* group more than the *low-risk* group, the effective combined $R_0$ could be improved, with commensurate reduction in hospitalization and fatality rates, without increasing adverse effects on the economy, as the most economically active group would face no additional restrictions on distancing. While numerical simulations on sophisticated models have included many cases of this kind of enhanced social distancing for the elderly,[9] the contribution of this paper is a simple way to quickly assess the effects of design parameters, such as cut-off age or distancing targets for high- and low-risk groups. The standard SIR model with two-subpopulations is used. Exact or approximate *analytical* solutions are obtained that can easily asses the effects of decision variables on outcomes such as hospitalization and critical care needs or fatalities. It is emphasized that the intent of the paper is not to advocate any particular policy. Rather, it purports to provide shortcut tools that both scientists and public officials can easily use in rapid what-if analysis of related ideas.

## 2. Background and context

The standard SIR model[12-15] comprises the basic equations

$$\frac{ds}{dt} = -\beta si \quad (1)$$

$$\frac{di}{dt} = \beta si - \gamma i = \beta \left(s - \frac{1}{R_0}\right) i \quad (2)$$

for the fractions, $s(t)$ and $i(t)$, referring to susceptible to a spreading infection and infectious, respectively, in a population of constant size. The third fraction,

$$r(t) = 1 - s(t) - i(t) \quad (3)$$

refers to individuals removed from the infectious group, by either recovery or death. Clearly, then,

$$\frac{dr}{dt} = \gamma i \quad (4)$$

where $r(0) = 0, i(0) = \epsilon, s(0) = 1 - \epsilon$; $\beta$ is the infection *spread* rate; $\gamma$ is the *removal* rate from the infectious group; and $R_0 \stackrel{\text{def}}{=} \beta/\gamma$ is the *basic reproductive ratio*. NPI aims at designing interventions, such as closing schools, churches, and social venues, workplace distancing, or quarantine, in order that the resulting universal $R_0$ – as uncertain as $R_0$ may be – best achieve desired outcomes.[1,16]

Despite its utter simplicity, the SIR model provides valuable insights for the course of an epidemic,[15] which are relevant to the two-tier strategy to be presented in the next section. In summary, eqn. (2) implies that an epidemic does not occur iff

$$R_0 < 1 \quad (5)$$

For $R_0 > 1$, eqns. (1)-(4) imply that (a) the infection spreads and an epidemic occurs (b) the infectious fraction, $i$, initially rises exponentially at rate $\beta - \gamma > 0$, (c) peaks at time $t^*$ with

$$s(t^*) = \frac{1}{R_0} < 1 \quad (6)$$

and (d) asymptotically goes to 0, achieving *herd immunity* by leaving the remaining uninfected fraction of the population, $s(\infty)$, at

$$s(\infty) = 1 - r(\infty) = -\frac{W(-e^{-R_0 R_0})}{R_0} \quad (7)$$

(Figure 1) where $W(z)$ is the *Lambert function*.[17,18]* It is eqn. (7) that underlies the stark predictions for anticipated fatalities from COVID-19 and the challenges faced by NPI in the absence of a vaccine, as already discussed in the Introduction.

Responses to the epidemic clearly can involve sophisticated policies that adjust $R_0$ over time, over geographical areas, or over different population groups. What follows is not an attempt to replace these indispensable tools. Rather, it is an attempt to present shortcut methods for rapid what-if analysis.

* *Interestingly, the analytical solution of the algebraic equation $1 - r(\infty) = exp(-r(\infty)R_0)$ for the SIR model in terms of the Lambert function $W(z)$, eqn. (4), pointed out as early as 1996,[17] if not earlier, may have escaped the attention of most literature in this field.[15]*

## 3. Flattening the curve: Benefits and compromises

As already discussed, "Flattening the Curve" by effecting a low $R_0$ (well below 2.5, near or below 1) is a universal approach to mitigating the total number of fatalities throughout the epidemic (Figure 1) as well as the number of infected patients in need of hospitalization or critical care at any time. It can be shown (*Appendix B*) that the appropriate $R_0$, given a maximum infectious fraction, $i^*$, is given analytically by

$$R_0 = \frac{W_{-1}\left(\frac{i^*-1}{e}\right)}{i^*-1} \quad (8)$$

where $W_{-1}$ is the lower branch of the Lambert function[18] with values less than $-1$.

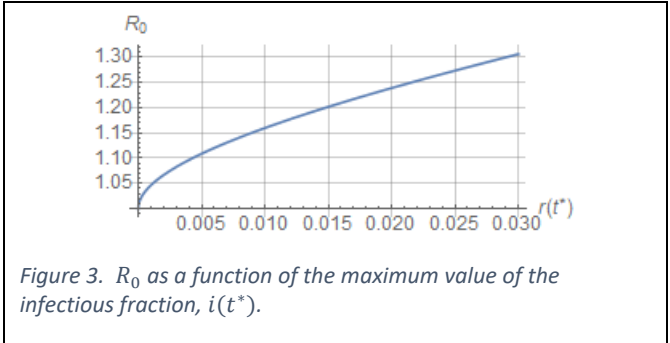

*Figure 3. $R_0$ as a function of the maximum value of the infectious fraction, $i(t^*)$.*

Reducing the universal $R_0$ value, however, inevitably prolongs the time, $t^*$, to the lowered peak of the infectious fraction, $r(t^*)$, hence the duration of the epidemic as well. This can be roughly captured by the time, $\tilde{t}$, needed for $i$ to reach an inflection, according to the equation

$$\frac{\tilde{t}_1}{\tilde{t}_2} \approx \frac{R_{0,2}-1}{R_{0,1}-1} \quad (9)$$

(*Appendix B*).

Eqn. (8) also suggests that while the time to peak decreases as $i(0) = \epsilon$ increases, the peak value of $i(t^*)$ remains approximately constant over a wide range of values of $\epsilon$, given a fixed value of $R_0$ as illustrated in Figure 4. This has obvious implications for designing strategies to flatten the curve.

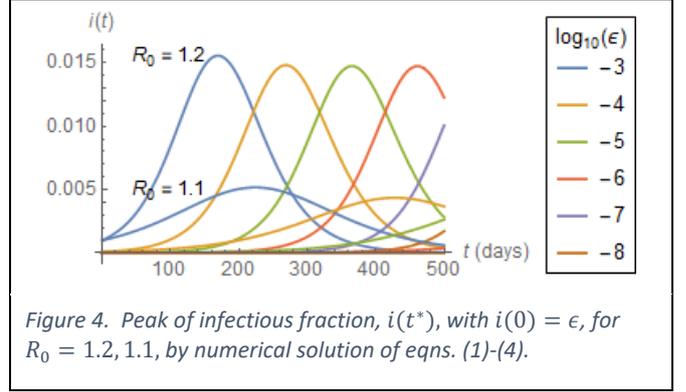

*Figure 4. Peak of infectious fraction, $i(t^*)$, with $i(0) = \epsilon$, for $R_0 = 1.2, 1.1$, by numerical solution of eqns. (1)-(4).*

## 4. Two-tier SIR model

In a population consisting of *high-risk* and *low-risk* groups a number of possibilities exist for adjusting binary contacts between groups. Assuming different rates of infection in contacts between the groups, the SRI model becomes

$$\frac{ds_H}{dt} = -\phi\beta s_H(\psi i_H + i_L) \quad (10)$$

$$\frac{ds_L}{dt} = -\beta s_L(\psi i_H + i_L) \quad (11)$$

$$\frac{di_H}{dt} = \phi\beta s_H(\psi i_H + i_L) - \gamma_H i_H \quad (12)$$

$$\frac{di_L}{dt} = \beta s_L(\psi i_H + i_L) - \gamma_L i_L \quad (13)$$

$$\frac{dr_H}{dt} = \gamma_H i_H \quad (14)$$

$$\frac{dr_L}{dt} = \gamma_L i_L \quad (15)$$

with initial conditions $s_H(0) = w_H - \epsilon_H$, $s_L(0) = w_L - \epsilon_L$, $i_H(0) = \epsilon_H$, $i_L(0) = \epsilon_L$. The positive parameters $\phi \leq 1$ and $\psi \leq 1$ capture the additional measures taken by the high-risk group, to avoid contacts.

Eqns. (12) and (13) imply that $i(t)$ declines after $t^*$ if

$$R_{0L}s_L(t^*) + R_{0H}\phi s_H(t^*) < 1/\psi \quad (16)$$

where $R_{0L} \stackrel{\text{def}}{=} \frac{\beta_L}{\gamma_L}, R_{0H} \stackrel{\text{def}}{=} \frac{\beta_H}{\gamma_H}$ (*Appendix A*). If $\phi = \psi = 1$, the herd immunity inequality, eqn. (6), is recovered.

In the rest of the paper we assume $\phi = \psi$ and $\gamma_H = \gamma_L$.

## 5. Reducing critical care needs and fatalities

Dividing eqn. (10) by eqn. (11) and integrating implies that

$$\frac{s_H(t)}{s_H(0)} = \left(\frac{s_L(t)}{s_L(0)}\right)^\phi \quad (17)$$

For $\phi = 1$ eqn. (17) simply confirms that both groups are infected at the same rate. In the extreme, if unrealistic case of $\phi = 0$ (quarantine of the high-risk group) it would follow from eqn. (17) that $s_H(t) = s_H(0)$, suggesting that only the low-risk group would be infected, with much lower needs for hospitalization and fatality rates.

For practically more interesting intermediate values of $\phi$, realistically closer to 1, it turns out (as explained next) that both fatalities and critical care needs are reduced, as shown in Figure 5.

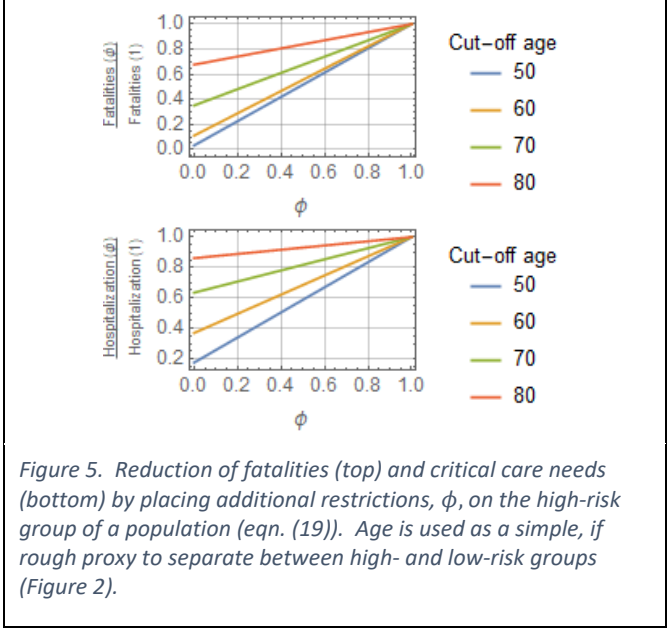

*Figure 5. Reduction of fatalities (top) and critical care needs (bottom) by placing additional restrictions, $\phi$, on the high-risk group of a population (eqn. (19)). Age is used as a simple, if rough proxy to separate between high- and low-risk groups (Figure 2).*

To generate the results in Figure 5, it is first assumed that the population fraction of low-risk infected by the end of the epidemic,
$$r_L(\infty) = s_L(0) - s_L(\infty) \leq s_L(0),$$
is set at an upper bound, to cap fatalities, given a level of restrictions $R_0$ on that group; whereas the high-risk group can manage additional restrictions $\phi R_0$, to experience a corresponding population fraction of infected
$$r_H(\infty) = s_H(0) - s_H(\infty) \leq s_H(0).$$
Then, the fatality fraction, $f$, for the entire population by the end of the epidemic will be
$$f = r_H(\infty) f_H + r_L(\infty) f_L \quad (18)$$

It can be shown (*Appendix C*) that the ratio of the total fatality fraction $f$ for $\phi < 1$ and for $\phi = 1$ is approximately independent of the selected value of $r_L(\infty)$ for reasonably small values of $r_L(\infty)$, e.g. below $0.1 - 0.2$, as
$$\frac{f(\phi)}{f(1)} \approx \frac{\frac{s_H(0)}{1-s_H(0)}\left(\frac{f_H}{f_L}\right)\phi+1}{\frac{s_H(0)}{1-s_H(0)}\left(\frac{f_H}{f_L}\right)+1} \quad (19)$$

A similar expression can be derived in the same way for hospitalization and critical care needs. In both cases, derivation relies in the simple approximation of eqn. (17)
$$\frac{r_H(\infty)}{s_H(0)} \approx \phi \frac{r_L(\infty)}{s_L(0)} \quad (20)$$

Note that the results in this section have been derived with minimal assumptions for the underlying model used.

## 6. Total fatalities for the two-tier strategy

Similar to eqn. (7), the total fraction of infected through the epidemic for each of the high- and low-risk groups can be shown (*Appendix D*) to be
$$\frac{r_L(\infty)}{s_L(0)} \approx 1 + \frac{W[-e^{-z}z]}{z} \quad (21)$$
$$z \stackrel{\text{def}}{=} R_0\bigl(\phi^2 s_H(0) + s_L(0)\bigr)$$
$$\frac{r_H(\infty)}{s_H(0)} = 1 - \left(1 - \frac{r_L(\infty)}{s_L(0)}\right)^\phi \approx \phi\left(1 + \frac{W[-e^{-z}z]}{z}\right) \quad (22)$$

Note that for $\phi = 1$, eqns. (21) and (22) trivially recover eqn. (7), with $\frac{r_L(\infty)}{s_L(0)} = \frac{r_H(\infty)}{s_H(0)} = \frac{r(\infty)}{1-\epsilon} = 1 + \frac{W(-e^{-R_0}R_0)}{R_0}$. Figure 6 illustrates eqns. (21) and (22).

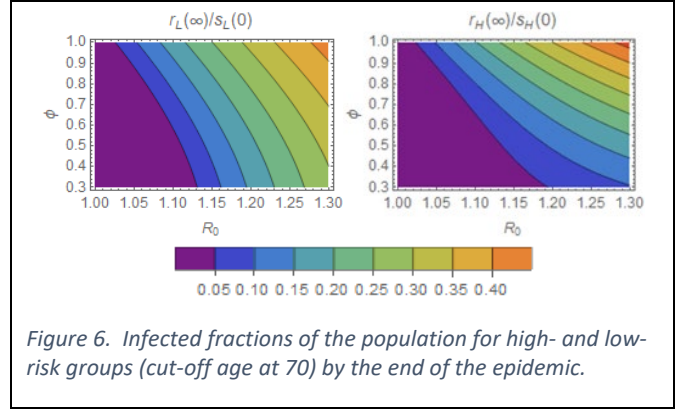

*Figure 6. Infected fractions of the population for high- and low-risk groups (cut-off age at 70) by the end of the epidemic.*

Finally, combined with eqn. (18), eqns. (21) and (22) yield the number of fatalities for the US shown in Figure 7.

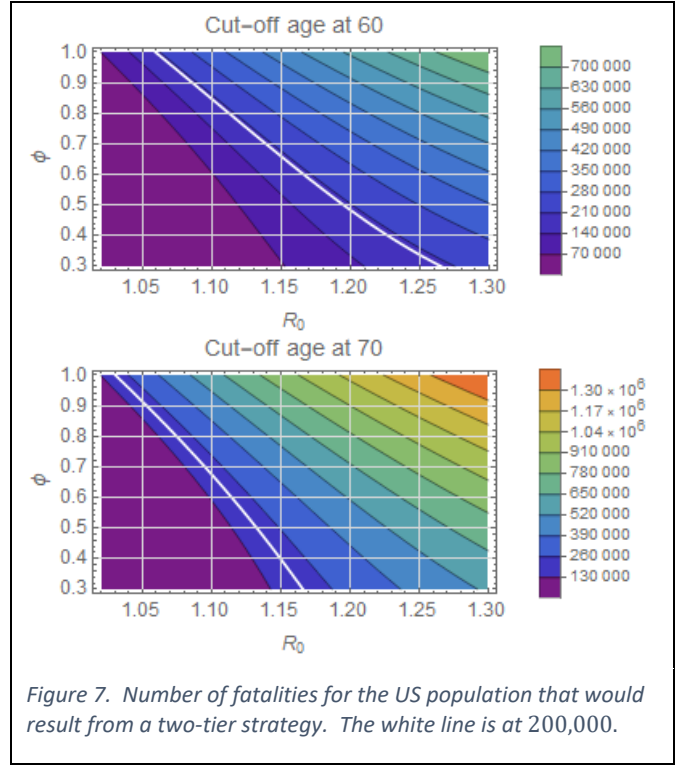

*Figure 7. Number of fatalities for the US population that would result from a two-tier strategy. The white line is at 200,000.*

## 7. Discussion

Simple analytical formulas for a two-tier NPI strategy to "flatten the curve" were collected. The formulas are based on a simple two-tier model of the SIR variety, involving high- and low-risk groups, and are meant to be used for rapid answers to what-if questions, rather than to provide accurate numbers. For the latter, numerical simulation remains the tool of choice. Taming the COVID-19 epidemic is a multifaceted problem that will require diverse resources and approaches. It is hoped that this text will contribute to that effort.

## 8. Acknowledgements

All computations were done in *Mathematica*, available at the University of Houston. Sharing of teaching material about the SIR model on Github by Jeff Kantor of Notre Dame is also gratefully acknowledged.

*Appendix A. Proof of eqn. (16)*

Eqns. (12) and (13) imply

$$\begin{bmatrix} di_H/dt \\ di_L/dt \end{bmatrix} = \underbrace{\begin{bmatrix} \phi\beta\psi s_H - \gamma_H & \phi\beta s_H \\ \beta\psi s_L & \beta\psi s_L - \gamma_L \end{bmatrix}}_{\mathbf{A}} \begin{bmatrix} i_H \\ i_L \end{bmatrix} \quad (23)$$

The eigenvalues of the matrix **A** are in the left half-plane iff

$$\det(\mathbf{A}) > 0 \quad (24)$$

which yields eqn. (16), and

$$\text{trace}(\mathbf{A}) < 0, \quad (25)$$

which is guaranteed by eqn. (16) if $\gamma_H \approx \gamma_L$.

*Appendix B. Proof of eqns. (8) and (9)*

Standard analysis of eqns. (1) and (2) proceeds as follows:

$$\frac{ds}{di} = -\frac{\beta s}{\beta s - \gamma} \Rightarrow i = 1 - s + \frac{1}{R_0}(\ln s - \ln(1-\epsilon))$$

If $i$ is maximized at $t^*$, then

$$i'(t^*) = 0 \Rightarrow s(t^*) = \frac{1}{R_0} \Rightarrow$$

$$i(t^*) \stackrel{\text{def}}{=} i^* = 1 - \frac{1}{R_0} - \frac{\ln R_0 + \ln(1-\epsilon)}{R_0} \Rightarrow$$

$$\boxed{i^* \approx 1 - \frac{1}{R_0} - \frac{\ln R_0}{R_0}} \Rightarrow \quad (26)$$

$$e^{R_0(i^*-1)}R_0(i^*-1) = \frac{i^*-1}{e}$$

Because $-\frac{1}{e} \leq \frac{i^*-1}{e} < 1$, eqn. (8) follows immediately.

The infectious fraction reaches an inflection point at $\tilde{t}$ iff

$$i''(\tilde{t}) = (\beta\tilde{s} - \gamma\tilde{\imath})' = 0 \Rightarrow \tilde{\imath} \approx \left(1 - \frac{1}{R_0}\right)^2$$

for $s(t) \approx 1$.

For $s(t) \approx 1$ and $i(0) = \epsilon \approx 0$, eqn. (2) implies

$$\tilde{\imath} \approx \epsilon e^{(\beta-\gamma)\tilde{t}} \Rightarrow \tilde{t} \approx \frac{1}{(\beta-\gamma)}(\ln(\tilde{\imath}) - -\ln(\epsilon)) \Rightarrow$$

$$\Rightarrow \frac{t_1^*}{t_2^*} \approx \frac{\beta_2 - \gamma}{\beta_1 - \gamma}\left(\frac{-\ln(\epsilon)}{-\ln(\epsilon)}\right) = \frac{R_{0,2} - 1}{R_{0,1} - 1}$$

which is eqn. (9).

---

*Appendix C. Proof of eqn. (19)*

Given a reasonable range of values for $\frac{s_L(\infty)}{s_L(0)} = 1 - \frac{r_L(\infty)}{s_L(0)}$ the approximation

$$\left(\frac{s_L(\infty)}{s_L(0)}\right)^\phi = \left(1 - \frac{r_L(\infty)}{s_L(0)}\right)^\phi \approx 1 - \phi\frac{r_L(\infty)}{s_L(0)} \quad (27)$$

is accurate, as indicated by the following figure.

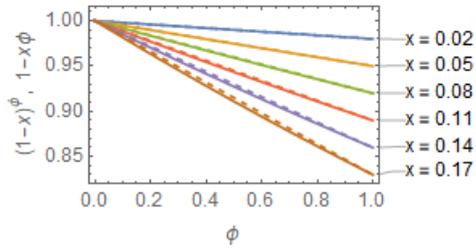

Therefore, eqns. (18) and (17) yield

$$\frac{f(\phi)}{f(1)} = \frac{\left(s_H(0) - s_H(0)\left(1 - \frac{r_L(\infty)}{s_L(0)}\right)^\phi\right)f_H + (s_L(0) - s_L(\infty))f_L}{\left(s_H(0) - s_H(0)\left(1 - \frac{r_L(\infty)}{s_L(0)}\right)\right)f_H + (s_L(0) - s_L(\infty))f_L} \approx$$

$$\approx \frac{\left(s_H(0) - s_H(0)\left(1 - \phi\frac{r_L(\infty)}{s_L(0)}\right)\right)f_H + (s_L(0) - s_L(\infty))f_L}{\left(s_H(0) - s_H(0)\left(1 - \frac{r_L(\infty)}{s_L(0)}\right)\right)f_H + (s_L(0) - s_L(\infty))f_L} =$$

$$= \frac{\frac{s_H(0)f_H}{s_L(0)f_L}\phi + 1}{\frac{s_H(0)f_H}{s_L(0)f_L} + 1}$$

which is eqn. (18).

---

*Appendix D. Proof of eqns. (21) and (22)*

Eqns. (11), (14), and (15) imply

$$\frac{ds_L}{d(\phi r_H + r_L)} = -\frac{\beta}{\gamma}s_L \Rightarrow$$

$$\Rightarrow s_L(\infty) = s_L(0)e^{-R_0(\phi r_H(\infty) + r_L(\infty))} \Rightarrow$$

$$\Rightarrow s_L(0) - r_L(\infty) = s_L(0)e^{-R_0(\phi r_H(\infty) + r_L(\infty))}$$

and, by eqn. (20),

$$\Rightarrow s_L(0) - r_L(\infty) = s_L(0)e^{-R_0\left(\phi^2\frac{s_H(0)}{s_L(0)} + 1\right)r_L(\infty)}$$

which implies eqn. (21), and, by eqn. (20) again, eqn. (22).